\documentclass[12pt]{article}
\usepackage{amsmath,amssymb}
\usepackage[dvips]{graphicx}
\usepackage{hyperref}
\usepackage{cite}
\usepackage{xcolor}

\makeatletter \@addtoreset{equation}{section} \makeatother
\makeatletter \@addtoreset{figure}{section} \makeatother

\makeatletter \@addtoreset{equation}{section} \makeatother
\makeatletter \@addtoreset{figure}{section} \makeatother

\def\CA{{\cal A}}\def\CD{{\cal D}}
\def\CF{{\cal F}}
\def\CI{{\cal I}}\def\CJ{{\cal J}}
\def\CL{{\cal L}}
\def\CN{{\cal N}}\def\CO{{\cal O}}
\def\CQ{{\cal Q}}
\def\CT{{\cal T}}\def\CV{{\cal V}}

\def\a{\alpha}\def\b{\beta}\def\g{\gamma}
\def\d{\delta}\def\e{\epsilon}
\def\th{\theta}
\def\l{\lambda}
\def\m{\mu}\def\n{\nu}
\def\r{\rho}\def\s{\sigma}

\def\G{\Gamma}
\def\Th{\Theta}

\def\vare{\varepsilon}

\newcommand{\bCA}{\bar\CA}

\begin{document}

%
\begin{titlepage}
\vfill
\begin{flushright}
{\tt\normalsize DAMTP-2013-25}\\
{\tt\normalsize EFI-13-6}\\
{\tt\normalsize PUPT-2447}\\
\end{flushright}
\vfill

\medskip
\begin{center}
\LARGE \bf
3d Chern-Simons Theory from M5-branes
\end{center}

\vfill

\begin{center}
\def\thefootnote{\fnsymbol{footnote}}

Sungjay Lee$^\diamondsuit$ and
Masahito Yamazaki$^\spadesuit$

\end{center}

\begin{flushleft}\small

\vskip 5mm

\centerline{$^\diamondsuit$ DAMTP, University of Cambridge, Cambridge, CB3 0WA, UK}

\vskip 5mm

\centerline{$^\diamondsuit$ Enrico Fermi Institute, University of Chicago, Chicago, IL 60637, USA}

\vskip 5mm

\centerline{$^\spadesuit$ Princeton Center for Theoretical Science, Princeton University,
Princeton, NJ 08544, USA}
\end{flushleft}

\vfill

\thispagestyle{empty}
\setcounter{tocdepth}{2}


\begin{center}
{\bfseries Abstract}
\end{center}

We study 5d $\mathcal{N}=2$ maximally supersymmetric
Yang-Mills theory with a gauge group $G$ on $S^2\times M_3$,
where $M_3$ is a 3-manifold. By explicit localization computation we
show that the path-integral of the 5d ${\cal N}=2$ theory reduces to
that of the 3d $G_{\mathbb{C}}$ Chern-Simons theory on $M_3$,
where $G_{\mathbb{C}}$ is the complexification of $G$.
This gives a direct derivation of the
appearance of the Chern-Simons theory from the
compactification of the 6d $(2,0)$ theory,
confirms the predictions from the
3d/3d correspondence for $G=SU(N)$,
and suggests the
generalization of the correspondence to
more general gauge groups.

\vfill

\end{titlepage}

\newpage
\setcounter{page}{1}

\tableofcontents

\parskip 0.1 cm

\section{Introduction}

We have learned over the past a few years that
there exists  beautiful correspondence (3d/3d correspondence) between
supersymmetric 3d $\CN=2$ theories and the geometry of 3-dimensional manifolds
\cite{Terashima:2011qi,Dimofte:2011ju,Cecotti:2011iy,Dimofte:2011py}.
(See also related earlier works \cite{Drukker:2010jp,Dimofte:2010tz,Hosomichi:2010vh})
\footnote{See also the generalization \cite{Terashima:2013fg} and
the related proposals in \cite{Terashima:2012cx,Yamazaki:2012cp}.}
The correspondence states that for a given 3-manifold $M_3$
there exists a corresponding 3d $\mathcal{N}=2$ theory
$\CT[M_3]$ defined from the compactification of the 6d $A_{N-1}$ $(2,0)$
theory on $M_3$, and that the $S^3$ partition function \cite{Kapustin:2009kz,
Jafferis:2010un,Hama:2010av} or the $S^1\times S^2$ index \cite{Kim:2009wb,
Imamura:2011su} of the 3d ${\cal N}=2$ theory $\CT[M_3]$ coincides with
the either full or holomorphic partition function of the $SL(N,\mathbb{C})$ Chern-Simons theory
on $M_3$.

While there have already been many works on this subject,
the identification of the 3d $\CN=2$ theory tends to rely on
indirect physical and mathematical arguments,
and there are often restrictions on the choices of $M_3$.
It is thus highly desirable to directly verify the claim
that the compactification of the 6d $(2,0)$
theory gives rise to the 3d $SL(N,\mathbb{C})$ Chern-Simons theory on $M_3$.

In this paper we study 5d $\CN=2$ maximally supersymmetric Yang-Mills (SYM)
theory with a gauge group $G$ on $S^2\times M$.
This is the dimensional reduction of the
6d $(2,0)$ theory on $S^1\times S^2\times M$.\footnote{For
non-simply-laced gauge groups
we need to turn on appropriate twists
along the $S^1$ direction.}
We show directly by supersymmetric localization
that the partition function of the 5d $\CN=2$ theory with
a gauge group $G$
reduces to the partition function of the
3d $G_{\mathbb{C}}$ Chern-Simons theory,
where $G_{\mathbb{C}}$ is the complexification of the
gauge group $G$:
\begin{align}
Z_{\textrm{5d $G$ SYM}}[S^2\times M_3] = Z_{\textrm{3d $G_{\mathbb{C}}$
 CS}}[M_3] \ .
\label{mainresult}
\end{align}

Let us denote the 5d gauge coupling constant by $g$,
and the radius of $S^2$ by $r$.
Both of these parameters are dimension-full,
and we find in the correspondence \eqref{mainresult}
that their dimension-less combination $\frac{r}{g^2}$
is identified with the
complexified level $t$ of the Chern-Simons theory by
\begin{align}
i\, t =\frac{8\pi^2 r}{g^2} \ .
\label{trelation}
\end{align}
The right hand side of \eqref{mainresult} is independent of
the size of the manifold $M_3$ since Chern-Simons theory
is a topological theory. We also find that there is no dependence
on the size of $M_3$ on the left.
The fact that the level $t$ is pure imaginary is consistent
with the result in \cite{Dimofte:2011py} where the
imaginary part of the level is identified as the chemical potential
of the 3d superconformal index.

Our computation  can be thought of a higher-dimensional lift of
the 2d localization on $S^2$ \cite{Benini:2012ui,Doroud:2012xw,Gomis:2012wy}
as well as the localization of the 4d twisted
$\CN=4$ Yang-Mills on $I\times M_3$ \cite{Witten:2010cx,Witten:2010zr}
(see appendix \ref{sec:CS} for the latter).
For $G=SU(N)$ we have $G_{\mathbb{C}}=SL(N, \mathbb{C})$
and our results can be regarded as the derivation of the
3d/3d correspondence existing in the literature, when the 3d $\CN=2$ theory on $S^2\times S^1$
is dimensionally
reduced along the $S^1$.\footnote{Our work is similar in spirit to the work of
\cite{Fukuda:2012jr,Kawano:2012up}, which discuss
5d SYM on a product of $S^3$ and a 2d Riemann surface
and obtained 2d $q$-deformed Yang-Mills theory.
This is consistent with the results of \cite{Gadde:2011ik}
and its reduction \cite{Nishioka:2011dq}.
}
While the precise formulation of the
3d/3d correspondence is currently unknown for
other gauge groups, our result \eqref{mainresult}
strongly suggests such generalizations.

\bigskip

The rest of this paper is organized as follows.
In section \ref{sec:N=2} we
introduce the Lagrangian and the supersymmetry (SUSY)
transformations of the 5d $\CN=2$ theory
on $S^2\times M_3$ partially topologically twisted
along the $M_3$.
In section \ref{sec:localization}
we add a $\CQ$-exact term to the Lagrangian and
show that classical saddle point configurations
reproduce the classical action of the complex Chern-Simons theory,
while the one-loop determinant is independent of the position of the
saddle point. In section \ref{sec:remarks} we conclude with some supplementary remarks on our results.

We also include three appendices. In appendix \ref{sec:CS}
we summarize the 3d $SL(N, \mathbb{C})$
Chern-Simons theory
and its relation with the 5d $\CN=2$ SYM and the 6d $(2,0)$ theory.
In appendix \ref{sec:N1SUSY}
we study 5d $\CN=1$ SYM on $S^2\times M_3$,
highlighting the differences from the similar analysis for
the 5d $\CN=2$ SYM in section \ref{sec:N=2}.
In appendix \ref{sec:S2} we review the
construction of 2d $\CN=(2,2)$ theory
with a twisted vector multiplet and charged twisted chiral multiplets.

\bigskip
\noindent
\underline{{\it Note Added:}}
During the preparation of this paper
we have been notified that Daniel~L. Jafferis and Clay Cordova
have been working on closely related projects,
and we have coordinated submission of our papers.
Towards the completion of this project we received the
paper \cite{Yagi}, which also discuss 5d $\CN=2$ SYM on $S^2\times M_3$
and obtained 3d TQFT closely related with the 3d Chern-Simons
theory. The computational methods, the Lagrangian
as well as the resulting expressions, however, seem to be different from ours.


\section{\texorpdfstring{5d $\CN=2$ SYM on $S^2\times M_3$}{5d N=2 SYM
 on S2 times M3}}\label{sec:N=2}

In this section we study the Lagrangian and the
supersymmetry transformations of the
5d $\CN=2$ supersymmetry on $S^2\times M_3$, partially topologically
twisted along $M_3$.\footnote{From now on, $\CN=1$ or $\CN=2$ refers to
the numbers of supercharges that theories under study preserves in the
flat space.} As will be argued shortly,
the supersymmetry algebra of the theory is given by
$SU(2|1)$ which contains $SU(2)$ charge generating
the isometries of $S^2$ and $U(1)_R$ R-symmetry charge.
Let us first present various facts which are consistent
to the supersymmetry algebra $SU(2|1)$.

This theory originates from the 6d $(2,0)$ theory on
$S^1 \times S^2 \times M_3$. Upon the compactification along $M_3$,
it is expected from the 3d/3d correspondence to have the superconformal index of a certain 3d $\CN = 2$ theory.
It is therefore natural to require the supersymmetry algebra
to be parameterized by the six-dimensional
Killing spinors on $S^1 \times S^2$ and constant spinors on $M_3$.
The reduction along $S^1$ to obtain the 5d gauge theory breaks $Sp(4)$, the R-symmetry
group in the flat space, down to $SO(3)_R \times SO(2)_R$. This is because
the 6d Killing spinors depend on the circle. One also has to twist the 5d
theory with $SO(3)_R$ to admit constant spinors on the curved manifold $M_3$ ({\it cf.}\
\cite{Marcus:1995mq,Blau:1996bx}).

Thus, the Lorentz group and R-symmetry group of the 5d gauge
theory on $S^2\times M_3$ should be $SU(2)$ and $U(1)$.
One can further show that the quarter of sixteen supercharges can survive after
the topological twisting. These facts are consistent to $SU(2|1)$.
In fact, this $SU(2|1)$ supersymmetry
can be identified with the supersymmetry
of the 2d $\CN=(2,2)$ theory
on $S^2$ \cite{Benini:2012ui,Doroud:2012xw} obtained from the dimensional
reduction along the curved manifold $M_3$.

As a side remark, one might naively imagine the following two-step
procedure: First construct a {\it physical} 5d gauge
theory on $S^2\times\mathbb{R}^3$. Then,
topologically twist the theory along $M_3$, when we replace
$\mathbb{R}^3$ by a curved 3-manifold $M_3$.
By a {\it physical} theory, we mean a theory
which does not involve any topological twisting and preserves at least
minimal supersymmetry, i.e., eight supercharges.
We present explicitly in appendix B a {\it physical} theory
on $S^2\times \mathbb{R}^3$ which is different to what we construct in section 2.1.
However, as soon as we place the theory on $M_3$ with a suitable twist, the R-symmetry
is completely broken. It implies that such a physical theory has no chance to
preserve $SU(2|1)$ after the topological twisting, which does not fit with the 3d/3d correspondence.



Without further ado, we present the explicit Lagrangian for the
5d $\CN=2$ SYM on $S^2\times M_3$ preserving
$SU(2|1)$ supersymmetry.

\subsection{Lagrangian}

Let us first summarize our conventions.
We use the indices $M, N, \cdots$ ($A, B, \cdots$) for the spacetime (internal space)
indices which runs from $1$ to $5$, while $I, J, \cdots$ for $Sp(4)$ R-symmetry indices.
We use the following representations for the five-dimensional gamma matrices $\G^M$ for the spacetime,
and $\hat\G^A$ for the internal space:
\begin{align}
  \G^m & = \g^m \otimes {\bf 1}_2 \qquad (m=1,2)\nonumber \ , \\
  \G^\m & = \g^3 \otimes \g^\m \qquad (\m=1,2,3) \nonumber \ , \\
  \hat \G^\m & = \g^\m \otimes \g^3 \qquad (\m=1,2,3) \nonumber \ , \\
  \hat \G^i & = {\bf 1}_2 \otimes \g^{i-3} \qquad (i=4,5) \ ,
\end{align}
where $\g^m = ( \tau^1 , \tau^2)$, $\g^\mu = (\tau^1,\tau^2,\tau^3)$
and $\tau_i$ are Pauli matrices. The charge conjugation operator $C$ and
the $Sp(4)_R$ invariant tensor $\hat C^{IJ}$ are given by
\begin{align}
  C & = \left( \tau^1 \right)^{ab} \otimes \e^{\dot a \dot b} \nonumber
 \ ,\\
  \hat C & = \e^{\a \b} \otimes \left( \tau^1 \right)^{\dot \a \dot \b}\ ,
\end{align}
where $\CI = (a,\dot a)$ ($I=(\a,\dot \a)$) denote the five-dimensional spinor
indices ($Sp(4)$ R-symmetry indices). Each of these indices $a$, $\dot a$,
$\a$ and $\dot \a$ is raised and lowered by the antisymmetric tensor
$\e^{ab}$, $\e^{\dot a \dot a}$, $\e^{\a\b}$ and $\e^{\dot\a \dot \b}$ with $\e^{12}=-\e_{12}=1$.
Our convention for bilinear of 5-dimensional spinors is
\begin{align}
  \vare \l = - \vare_{\CI} C^{\CI \CJ} \l_{\CJ}\ ,
  \qquad \vare \G^M \l = - \vare_{\CI} (C\G^M)^{\CI \CJ} \l_{\CJ}\ ,
  \qquad \text{etc.}
\end{align}

\paragraph{On Flat $\mathbb{R}^5$}
Let us begin with the maximally supersymmetric Yang-Mills theory in flat $\mathbb{R}^5$.
The $\CN=2$ vector multiplet contains a gauge field
$A_M$, scalar fields $\phi^A$ in ${\bf 5}$ of $Sp(4)$, and gaugino fields $\l_I$.
The Lagrangian is given by
\begin{align}
  \CL_{\mathbb{R}^5} & =  \frac{1}{g^2} \text{tr} \bigg[ \frac14 F_{MN} F^{MN} + \frac12(D_M \phi_A )^2
  + \frac i2 \l_I  {\hat C}^{IJ} \G^M D_M \l_J -  \frac14 [\phi_A,\phi_B]^2
  \nonumber \\ & \qquad \
  - \frac i2 \l_I (\hat C \hat \G^A)^{IJ} [ \phi^A, \l_J ]  \bigg] \ .
  \label{flatLag}
\end{align}
This is invariant under the following on-shell SUSY transformation
rules\footnote{This can be derived by the dimensional reduction of the
on-shell SUSY transformation for the 10d $\CN=1$ SYM theory.}
%
\begin{align}
  \d A_M & = i  \vare_I \hat C^{IJ} \G_M \l_J \nonumber  \ ,\\
  \d \phi_A & =  \vare_I (\hat C \hat \G_A )^{IJ} \l_J \nonumber  \ ,\\
  \d \l_I & = - \frac12 \G^{MN} \vare_I F_{MN} + i \G^M (\hat \G^A \vare)_I D_M \phi_A
  + \frac i2 (\hat \G^{AB} \vare)_I \big[ \phi_A, \phi_B \big]  \ ,
\end{align}
where we used
\begin{align}
F_{MN}&=\partial_M A_N-\partial_N A_M-i[A_M, A_N] \ , \nonumber\\
D_M \sigma&=\partial_M \sigma-i [A_M, \sigma] \ .
\end{align}

\paragraph{On-Shell SUSY on $S^2\times M_3$} From the 3d/3d correspondence we expect
that the supersymmetry parameters $\vare_I$ for the theory on
$S^2\times M_3$ should be a Killing spinor on $S^2$ and a constant spinor on $M_3$ satisfying
\begin{align}
  \nabla_m \vare_I = - \frac{i}{2r}\G_m \G_{12} \vare_I \ , \qquad \nabla_\mu \vare_I = 0\ .
\end{align}
In order to have a constant spinor on a curved three-manifold $M_3$, one suitably twists
the theory with $SO(3)_R$ subgroup of $Sp(4)$ ({\it cf.}
\cite{Marcus:1995mq,Blau:1996bx}).
Let us denote by $SO(3)_\text{twist}$ the diagonal subgroup of the $SO(3)$
local Lorentz group on $M_3$ and the $SO(3)_R$.
The leftover
$SO(2)_R$ is then identified as the $U(1)_R$ R-symmetry of the $SU(2|1)$
supersymmetry algebra on $S^2$.
Under the symmetry group $SO(3)_\text{twist} \times U(1)_R$, various fields can be decomposed as follows
\begin{eqnarray}
  \begin{array}{rcccl}
    A_M & : & {\bf 1}_{0} \oplus {\bf 3}_0 & \equiv & A_m \oplus A_\mu \ , \vspace*{0.3cm} \\
    \l_I  & : & {\bf 1}_{\pm 1} \oplus {\bf 3}_{\pm 1} & \equiv & \big( \l, \bar \l \big) \oplus \left(
    \psi^\mu, \bar \psi^\mu \right) \ , \vspace*{0.3cm} \\
    \phi^A & : &  {\bf 1}_{\pm 2} \oplus {\bf 3}_{0  }\, \  & \equiv & \varphi_{\pm} \oplus
    \phi^{\mu} \ ,
  \end{array}
\end{eqnarray}
%
while the supersymmetry parameters can be decomposed as
\begin{align}
  \vare_I \ : \ {\bf 1}_{\pm 1} \oplus {\bf 3}_{\pm 1}\ .
\end{align}
The $SU(2|1)$ supersymmetry of our interest can be parameterized by the singlets $(\xi,\bar \xi)$
under the $SO(3)_\text{twist}$, which takes the following form
\begin{align}
  \big( \vare_{I} \big)_{a\dot a}= \frac i2 \e_{\dot a \a} \left( \xi_a \otimes\vare^+_{\dot \a}
  - (\g^3 \bar \xi)_a \otimes \vare^-_{\dot \a}\right)\ ,
  \label{4Q}
\end{align}
where $\xi$ and $\bar \xi$ satisfy the Killing spinor equation on the two-sphere
\begin{align}
  \nabla_m \xi = + \frac{1}{2r} \g_m \g^3 \xi \ , \qquad
  \nabla_m \bar \xi = - \frac{1}{2r} \g_m \g^3 \bar \xi\ ,
\end{align}
and
\begin{align}
  \vare^+ = \begin{pmatrix} 1 \\ 0 \end{pmatrix}\ , \qquad
  \vare^- = \begin{pmatrix} 0 \\ 1 \end{pmatrix}\ .
\end{align}
%

For later convenience, let us Define $\varphi_\pm = \phi^4 \mp i \phi^5$,
$\CA_\mu = A_\mu + i \phi_\m$ and
\begin{align}
  \e^{\dot a \a}\l_{a\dot a \a \dot \a} & \equiv i \left( \l_a \otimes \vare^+_{\dot \a}
  - (\g^3\bar\l)_a\otimes \vare^-_{\dot \a}\right)
  \nonumber\ ,  \\
  \left( \e\g^\mu \right)^{\dot a \a} \l_{a\dot a \a \dot \a} & \equiv  \left(
  \psi^\mu_a \otimes \vare^+_{\dot \a} - (\g^3\psi^\m)_a \otimes \vare^-_{\dot \a} \right)
  \ .
\end{align}
In terms of new variables, the on-shell SUSY variation of the fields on $S^2\times M_3$
are given by
\begin{align}
  \d A_m & = - \frac12 \left( \xi \g_m \bar \l + \bar \xi \g_m \l \right) \ , \nonumber \\
  \d \varphi_+ & = - \xi \g^3 \l \ , \nonumber \\
  \d \varphi_- & = + \bar \xi \g^3 \bar \l \ , \nonumber \\
  \d \l & =  \left( i F_{12} + i D^\m \phi_\m \right) \g^3\xi
  + i \g^m\g^3 D_m (\bar \xi \varphi_+)
  + \frac i2 \big[\varphi_- , \varphi_+ \big] \xi \ ,
  \label{tv}
\end{align}
and
\begin{align}
  \d \CA_\mu  = &- \xi \g_+  \bar \psi_\m - \bar \xi \g_- \psi_\m \ , \nonumber \\
  \d \psi_\m  = & -i \g^m\g^3 \xi F_{m\m} + \g^m \xi D_m \phi_\m
  + D_\m \varphi_+ \bar \xi + \g^3 \bar \xi [ \varphi_+ , \phi_\m ]  \nonumber \\
  &
  +  \left[  \frac12 \e_{\m\n\r} \left\{ F^{\n\r}  + i ( D^\n \phi^\r - D^\r \phi^\n )\g^3
  + i [\phi^\n,\phi^\r] \right\}  \right] \xi \ ,
  \label{tc}
\end{align}
%
where $\g_\pm = \frac12 ( 1 \pm \g^3)$. Note here that we add $1/r$-correction terms
in the SUSY variation rule for the fermion field $\l$. Our convention for bilinear of 2-dimensional spinors is $\bar \xi \l = \xi^a \l_a = - \xi_a \e^{ab} \l_b$,
$\bar \xi \g^m \l = - \xi_a (\e \g^m)^{ab} \l_b$ and so on.

The $SU(2|1)$ invariant Lagrangian on $S^2\times M_3$ is then given
by
\begin{align}
  \CL = \CL_{\mathbb{R}^5} + \frac{1}{4rg^2} \left( \CL_\text{CS}(\CA) - \CL_\text{CS}(\bar \CA)\right) + \frac{i}{2r g^2} \text{tr}\Big[\bar \l \g^3 \l \Big]\
  \label{n2ym}
\end{align}
with
\begin{align}
  \CL_\text{CS}(\CA) = \e^{\m\n\r} \text{tr} \Big[\CA_\m \partial_\n \CA_\r - \frac{2}{3}i
  \CA_\m \CA_\n \CA_\r \Big]\ .
\end{align}
As commented at the beginning of this section,
the Lagrangian of the physical (\textit{i.e.}, non-twisted) SYM on $S^2 \times \mathbb{R}^3$
constructed in appendix \ref{sec:N1SUSY}
can not be embedded into the above Lagrangian of our interest.

\paragraph{Off-Shell SUSY on $S^2\times M_3$}
In the next section, we shall use the localization method to reduce the
path integral of the 5d $\CN=2$ theory on $S^2\times M_3$
down to that of a 3d theory on $M_3$. We need the off-shell
supersymmetry for this purpose. Introducing an auxiliary field $D$ whose on-shell value becomes
\begin{align}
  D = - i D^\m \phi_\m\ ,
\end{align}
and three complex auxiliary fields $G_\m$ whose one-shell values are
\begin{align}
  G_\m = \frac12 \e_{\m\n\r} \bar{\CF}^{\n\r}\ ,
\end{align}
the off-shell SUSY transformation rules are given by
\begin{align}
  \d A_m & = - \frac12 \left( \xi \g_m \bar \l + \bar \xi \g_m \l \right) \ , \nonumber \\
  \d \varphi_+ & = - \xi \g^3 \l \ , \nonumber \\
  \d \varphi_- & = + \bar \xi \g^3 \bar \l \ , \nonumber \\
  \d \l & =  \left( i F_{12} - D \right) \g^3\xi
  + i \g^m\g^3 D_m (\bar \xi \varphi_+)
  + i \big[\varphi_- , \varphi_+ \big] \xi \ , \nonumber \\
  \d D & = \frac i2 \bigg( D_m ( \xi \g^m\g^3 \bar \l - \bar \xi \g^m\g^3 \l  ) + [\varphi_+ ,
  \bar \xi \bar \l ] + [ \varphi_- , \xi \l ] \bigg) \ ,
  \label{offshelltv}
\end{align}
and
\begin{align}
  \d \CA_\mu  = &- \xi \g_+  \bar \psi_\m - \bar \xi \g_- \psi_\m \ , \nonumber \\
  \d \psi_\m  = & -i \g^m\g^3 \xi F_{m\m} + \g^m \xi D_m \phi_\m
  + D_\m \varphi_+ \bar \xi + \g^3 \bar \xi [ \varphi_+ , \phi_\m ]
  +  \big( {\bar G}_\mu \g_+  +  G_\mu \g_- \big) \xi \ , \nonumber \\
  \d G_\m  = & + i \xi \g^m \g_+ D_m \bar \psi_\m + \xi \g_- D_\m \bar \l
  + i \bar \xi \g^\m \g_- D_m \psi_\m
  +  \bar \xi \g_+ D_\m \l
  \nonumber \\ &
  + \big[ \phi_\m , \xi \g_- \bar \l +  \bar \xi \g_+ \l \big]
  + i \big[ \varphi_+ , \bar \xi \g_+ \bar \psi_\m \big]
  +  i \big[ \varphi_- , \xi \g_- \psi_\m \big]\ .
  \label{offshelltc}
\end{align}
The commutator $[\d_\eta , \d_\vare]$ reads
\begin{align}
  \big[\d_1 , \d_2 \big] = - i \CL_v + \text{Gauge}(\g) + \text{Lorentz}(\Theta) + U(1)_R(\a)\ ,
\end{align}
where
\begin{align}
  \CL_v & = v^M \partial_M  =  2 \bar \xi_1 \g^m \xi_2 \partial_m  \ , \nonumber \\
  \g & = i v^M A_M +  2  \bar \xi_1 \g^3 \bar \xi_2 \varphi_+ -  2\xi_1 \g^3 \xi_2 \varphi_- \ ,
  \nonumber \\
  \Theta & = - \frac 2r \bar \xi_1 \xi_2 \nonumber \ , \\
  \a & = \frac{1}{r} \bar \xi_1 \g^3 \xi_2  \ .
  \label{Hoperator}
\end{align}
It is illustrative to rewrite the bosonic part of the on-shell Lagrangian (\ref{n2ym})
in the off-shell SUSY invariant form,
\begin{align}
  \CL_{\rm b} = \CL_{\rm tv} + \CL_{\rm tc} + \CL_{W} + \CL_{\bar W}\ ,
\end{align}
where
\begin{align}
  \CL_{\rm tv}  = & \text{tr}\Big[ + \frac12 (F_{12})^2 + \frac12 D_m \varphi_+ D_m \varphi_- +
  \frac18 \big[\varphi_+, \varphi_- \big]^2 + \frac12 D^2 \Big]\ ,
  \nonumber \\
  \CL_{\rm tc}  = & \text{tr}\Big[ + \frac12 (F_{m\m})^2 + \frac12 (D_m \phi_\m)^2   +
  \frac12 D_\m \varphi_- D_\m \varphi_+ - \frac12 [\phi_\m, \varphi_+][\phi_\m,\varphi_-]
  \nonumber \\ & \hspace*{0.5cm}
  + \frac12 \bar G_\m  G^\m + i D (D^\m \phi_\m) \Big] \nonumber \ , \\
  \CL_W = & \text{tr} \Big[ - \frac i2 \frac{\partial}{\partial \CA^\m} W(\CA^\m) G^\m \Big]
  + i \frac{1}{2r} W(\CA^\m)
  \nonumber \ , \\
  \CL_{\bar W}  = & \text{tr}\Big[ - \frac i2 \frac{\partial}{\partial \bar{\CA}^\m} {\bar W}(\bar{\CA}^\m)
  \bar{G}^\m \Big]  + i \frac{1}{2r} {\bar W}(\bar{\CA}^\m) \ ,
  \label{twtheory}
\end{align}
with
\begin{align}
  W = - \frac{i}{2} \e^{\m\n\r} \text{tr} \Big[\CA_\m \partial_\n \CA_\r - \frac{2}{3}i
  \CA_\m \CA_\n \CA_\r \Big]\ , \qquad \frac{\partial}{\partial \CA^\m} W(\CA^\m) =
  -\frac i2 \e_{\m\n\r} \CF^{\n\r}\ .
\end{align}

Note that, in the language of $SU(2|1)$ supersymmetry algebra on $S^2$,
$(A_m, \varphi_\pm, \l,$ $D)$ transforms as a twisted vector multiplet while
$(\CA_\m, \psi_\m, G_\m)$ transform as twisted chiral multiplets. Moreover,
the bosonic Lagrangian terms $\CL_\text{tv}$, $\CL_\text{tc}$ and $\CL_W$ in (\ref{twtheory})
can be understood as kinetic terms for twisted vector and twisted chiral multiplet,
and twisted superpotential terms.
Readers are referred to the Appendix \ref{sec:S2} for details.

\section{Localization of the Path Integral}\label{sec:localization}

We show in this section that the path integral of 5d $\CN=2$ SYM
on $S^2\times M_3$ can be localized to the path integral of 3d $G_\mathbb{C}$
Chern-Simons theory on $M_3$ with purely imaginary Chern-Simons level. To this
end, we use the supersymmetric localization technique.

\subsection{Saddle Point Configurations}

\paragraph{Choice of Supercharge} In our localization scheme,
we choose a particular supersymmetry generator $\CQ$ in $SU(2|1)$
which is associated to the following
Killing spinors $\xi$ and $\bar \xi$ in \eqref{4Q}
\begin{align}
  \xi  = e^{+i\varphi/2}  \begin{pmatrix} \cos\frac{\th}{2} \\ \sin\frac{\th}{2} \end{pmatrix}\ ,
  \qquad
  \bar \xi =  e^{-i\varphi/2}  \begin{pmatrix} \sin\frac{\th}{2} \\		
  \cos\frac{\th}{2} \end{pmatrix}\  \ .
\end{align}
Here $(\th,\varphi)$ denote the polar coordinates on the two-sphere.
Given our choice of supercharge, we can show that
\begin{align}
  \d^2 = - i \CL_v + \text{Gauge}(\g) + \text{Lorentz}(\Theta) + U(1)_R(\a)\ ,
\end{align}
where
\begin{align}
  \CL_v & = v^M \partial_M  =  \bar \xi \g^m \xi \partial_m = - i \partial_\varphi \ , \nonumber \\
  \g & = i v^M A_M +   \bar \xi \g^3 \bar \xi \varphi_+ -  \xi \g^3 \xi \varphi_- =
  A_\varphi + e^{-i\varphi} \sin\th \varphi_+ - \e^{+i\varphi} \sin\th \varphi_- \ ,
  \nonumber \\
  \Theta & = - \frac 1r \bar \xi \xi = - \frac 1r \cos\th \nonumber \ , \\
  \a & = \frac{1}{2r} \bar \xi \g^3 \xi  = \frac{1}{2r} \ .
  \label{Hoperator}
\end{align}
Here we normalize the $U(1)_R$ charge so that $\l$ and $\psi^\m$
carry $+1$ R-charge while $\varphi_\pm$ carry $\pm 2$ R-charge.
For instance, the square of the supercharge $\CQ^2$ generates
the following infinitesimal transformation of the fermionic field $\l$
\begin{align}
  \d^2 \l = - i \CL_v \l + i [ \g , \l] + i \Theta \frac{\g^3}{2} \l + i \a \l\ .
\end{align}

\paragraph{Ghosts and BRST Symmetry} Once introducing the Faddeev-Popov
ghost field $c$ to fix a gauge, one can define the BRST transformation
as follows
\begin{gather}
  Q_B A_M = D_M c\ , \qquad  Q_B \phi^A = i [ c , \phi^A] \ , \qquad
  Q_B \l = i \{ c , \l \} \ ,
  \nonumber \\
  Q_B \psi = i \{ c , \psi \}\ , \qquad
  Q_B D = i [ c, D]\ , \qquad   Q_B c = \frac i2 [c,c] + a_0 \ ,
\end{gather}
where $a_0$ is a constant field. With the above rule, one can show that
\begin{align}
  Q_B^2 = \text{Gauge}(a_0) \ .
\end{align}
The fermionic symmetry  we use for the localization is $\hat Q = \CQ + Q_B$
which is required to satisfy
\begin{align}
  {\hat \CQ}^2 =  \left( \CQ + Q_B \right)^2 = -i \CL_v + \text{Gauge}(a_0)
  + \text{Lorentz}(\Theta) + U(1)_R(\a)\ .
\end{align}
It leads that the ghost field $c$ should transform under $\CQ$ as
\begin{align}
  \CQ c =   - \left(  i v^M A_M +
  \bar \xi \g^3 \bar \xi \varphi_+ -  \xi \g^3 \xi \varphi- \right)\ .
\end{align}
One can also show that the constant field $a_0$ should satisfy the relations below
\begin{align}
  \CQ a_0 = Q_B a_0 = 0\ .
\end{align}
The transformation rules for the anti-ghost fields are given by
\begin{gather}
  Q_B \bar c = B \ , \qquad Q_B B = i [a_0 , \bar c]\ ,
  \nonumber \\
  \CQ \bar c = 0 \ , \qquad \CQ B = - i \CL_v \bar c\ .
\end{gather}
To remove the constant modes of $c$ and $\bar c$, denoted by $c_0$ and $\bar c_0$,
we use the multiplets of constant fields satisfying
\begin{align}
  \hat Q \bar a_0 = \bar c_0\ , \qquad \hat Q \bar c_0 = i [ a_0 , \bar a_0 ] \  ,
  \qquad \hat Q B_0= c_0 \ , \qquad \hat Q c_0 = i [a_0, B_0] \ .
\end{align}

\paragraph{$\hat Q$-exact Deformations} We now turn into the choice of
$\hat Q$-exact deformation terms. It is convenient to choose them as follows
\begin{align}
  \CL_\text{def} \equiv \hat \CQ \CV =
  \hat \CQ \,  \bigg[ \d_{\bar \xi} \Big\{ \text{tr}\big[ \bar \l \g^3 \l
  + \bar \psi_\mu \g^3 \psi^\mu \big] \Big\} +  \text{tr}\big[ \bar c f + \bar c B_0
 + c \bar a_0 \big] \bigg]\ ,
  \label{Qexact}
\end{align}
which provides positive semi-definite bosonic terms once we impose the
standard reality conditions on the field variables. The deformed Lagrangian
also contains the gauge fixing term,
\begin{align}
  \CL_\text{def}= B f[A_M,\phi^A]+\cdots\ ,
\label{dfdsf}
\end{align}
where $f[A_M,\phi^A]$ is given by, for instance,
\begin{align}
  f[A_M,\phi^A] = i \partial^M A_M + \CL_v \big( \bar \xi \g^3 \bar \xi \varphi_+ + i v^M A_M \big) \ .
\label{f}
\end{align}

\paragraph{Saddle Point Configurations}
The next step is to find out the saddle point locus where
the given path integral can localize on. With the above choice
of supercharge $\hat Q$ and $\hat Q$-exact deformation terms,
we can describe the supersymmetric saddle point configurations,
satisfying the SUSY condition and the equations of motion from
the deformed action, as follows
\begin{align}
  \varphi_\pm = F_{12} = D = 0 \ , \qquad \CA_\m(x^M) = \CA_\m(x^\m) \ , \qquad
  G_\m  = 0 \ ,
\end{align}
and
\begin{align}
  f[A_M,\phi_A] = 0\ .
\end{align}
One can easily see from (\ref{dfdsf})
that the last condition arises from the equation of motion for the auxiliary field $B$.

Note that the saddle point condition for the ghost field, $\hat Q c = 0$, leads to
\begin{align}
  \frac i2 [c,c] + a_0 - \left(  i v^M \hat A_M +
  \bar \xi \g^3 \bar \xi \hat \varphi_+ -  \xi \g^3 \xi \hat \varphi- \right)= 0\ ,
\end{align}
where $\hat \varphi_\pm$ and $\hat A_M$ satisfy the condition for the saddle point configurations.
That is to say, it turns out that
\begin{align}
  a_0 = 0\ .
  \label{great}
\end{align}
As we will see below, this result (\ref{great}) implies that one-loop
determinant is independent of the saddle point configurations.

Plugging these saddle points into the
Lagrangian, we learn
that the path-integral of the 5d $\CN=2$ theory can be reduced to
the path-integral of a $G_{\mathbb{C}}$ Chern-Simons theory on $M_3$ up to a one-loop
factor:
\begin{align}
  Z_\text{5d} = \int \CD \CA_\m(x^\n) \CD \bar\CA_\m(x^\n) \ ,
  e^{- \frac{\pi r}{g^2}\int_{M_3} \ \big( \CL_\text{CS}(\CA) -
  \CL_\text{CS}(\bar \CA) \big)} Z_\text{one-loop}(\CA,\bar \CA)\ ,
\label{tmpresult}
\end{align}
where we sum over all the possible field-configurations $(\CA_\m,\bar \CA_\m)$
subject to the gauge-fixing condition (\ref{f}), i.e.,
\begin{align}
  f[A_\mu,\phi_\mu] = 0\ .
  \label{f2}
\end{align}
We will next compute the one-loop determinant $Z_\text{one-loop}(\CA,\bar \CA)$.

\subsection{One-loop Determinant}\label{subsec:index}

\paragraph{Review}
From our choice of the $\CQ$-exact deformation $\CL_\text{def}=t\hat Q \CV$ (\ref{Qexact}), we
can compute the one-loop determinant around the fixed-point configurations.
We will follow the standard procedure in the literature,  see
e.g. \cite{Pestun:2007rz,AtiyahLecture}.
In terms of a new set of path-integration variables, one can write
\begin{align}
  \CV = \begin{pmatrix} \hat Q X & \Psi \end{pmatrix}
  \begin{pmatrix} D_{00} & D_{01} \\ D_{10} & D_{11} \end{pmatrix}
  \begin{pmatrix} X \\ \hat Q \Psi \end{pmatrix}\ ,
\end{align}
where $X$ denote collectively bosonic variables while $\Psi$ denote fermionic variables.
One can then obtain the kinetic terms as follows
\begin{align}
  \CL_\text{def} = \CL_\text{b} + \CL_\text{f}\ ,
\end{align}
where
\begin{align}
  \CL_\text{b} & = \begin{pmatrix} X & \hat Q \Psi \end{pmatrix}
  \begin{pmatrix} H & \\ & 1 \end{pmatrix}
  \begin{pmatrix} D_{00} & D_{01} \\ D_{10} & D_{11} \end{pmatrix}
  \begin{pmatrix} X \\ \hat Q \Psi \end{pmatrix}\ ,
  \nonumber \\
  \CL_\text{f} & = \begin{pmatrix} \hat Q X  & \Psi \end{pmatrix}
  \begin{pmatrix} D_{00} & D_{01} \\ D_{10} & D_{11} \end{pmatrix}
  \begin{pmatrix} 1 & \\ & H \end{pmatrix}
  \begin{pmatrix} \hat Q X \\  \Psi \end{pmatrix} \ .
\end{align}
Note that $H=\hat Q^2$ commutes with the operators $D_{ij}$.
One can show formally the following relation
\begin{align}
  \left( \frac{\det K_f}{\det' K_b} \right)^2 = \frac{\det_\Psi H}{\det'_X H } =
  \frac{\det_{\text{Coker} D_{10}} H }{\det'_{\text{Ker} D_{10}} H}\ ,
\end{align}
where $K_f$ and $K_b$ denote the kinetic operators acting on fermionic and bosonic variables.
Here $\det'$ indicates that bosonic zero modes should be excluded from the determinant.
One can read off from the index below the ratio of determinants
\begin{align}
  \text{ind}\, D_{10}  = \text{Tr}_{\text{Ker} D_{10}} \left[ e^{-Ht} \right] - \text{Tr}_{\text{Coker} D_{10}}
  \left[ e^{-Ht} \right] \ .
\end{align}
Note that 5d $\CN=2$ supersymmetric YM on $S^2\times M_3$ has eighteen bosonic variables
$(A_m, \varphi_\pm, \CA_\m , \bar \CA_\m, D, G_\m,$ $\bar G_\m, B)$ and
eighteen fermionic variables $(\l, \bar \l, \psi_\m, \bar \psi_\m, c, \bar c)$.
To compute the index of the operator $D_{10}$, the new set of
path integration variable is chosen as
\begin{align}
  X_i = & \ (A_m , \ \varphi_+ , \ \CA_\m , \ \bar \CA_\m ; \bar a_0, B_0)
  \nonumber \\
  \Psi_i = & \ ( \bar \xi \g^3 \bar \l, \ c , \ \bar c, \ \bar\xi \g^m \psi_\m - \xi \g^m \bar \psi_\m )\ ,
\end{align}
and their descendants $\hat Q X$ and $\hat Q \Psi$.
Here $i$ runs from $1$ to $9$.

\paragraph{Index Computation}
Recall that the saddle point configurations can be described as
\begin{align}
  A_m = \varphi_\pm = 0 \ , \qquad \CA_\m = \hat \CA_\m(x^\m) \ .
  \label{saddlepointagain}
\end{align}
For simplicity, let us set $r=1$ from now on.
The index of the operator $D_{10}$ is
\begin{align}
  \text{ind}\, D_{10} = \text{Tr}_X \left[ e^{-t H} \right] - \text{Tr}_\Psi \left[e^{-t H} \right]\ ,
\end{align}
where the symbol Tr indicates a combined matrix (i.e., over the indices $i$ from $1$
to $9$
and group indices) and functional trace and
\begin{align}
  H = \hat Q^2 = - i \CL_v + \text{Lorentz}(\Th) + \text{Gauge}(a_0) + R_{U(1)}(\a) \ .
\end{align}
It acts on a field $\CO$ in adjoint representation as follows
\begin{align}
  e^{-tH} \CO(x^M) = h_{[\CO]}  \cdot e^{-ia_0(x^M) t} \CO(\tilde x^M) e^{+ia_0(x^M) t} \ , \qquad \tilde x =
  e^{ + t\partial_\varphi } x \ ,
\end{align}
where the factor $h_{[\CO]}$ encodes the action of $H$ on the vector and $U(1)_R$
indices of the field $\CO$. As explained in \cite{Hama:2012bg}, we ignore
terms containing constant fields $B_0$ and $\bar a_0$ from $\CV$ (\ref{Qexact}) in computing the index.
These constant fields are thus regarded as sitting in the kernel of $D_{10}$ leading to  a contribution $2$
to the index.

From the equations (\ref{Hoperator}) and (\ref{great}) one shows that the operator $H$ is
independent of the saddle point configurations (\ref{saddlepointagain}).
This means that the index of $D_{10}$ should be independent of $\CA_\m(x^\n)$ and $\bar \CA_\m(x^\n)$,
leading to {\it trivial one-loop determinant contributions} !
We can therefore conclude that
\begin{align}
  Z_\text{one-loop} (\CA,\bar \CA) = 1\ .
  \label{result5}
\end{align}

Let us then compute the index just for the completeness, although it
is irrelevant to the conclusion.
Massage the trace of $e^{-Ht}$ over $X$ fields into
\begin{align}
  \text{Tr}_{X_i} \left[ e^{-t H} \right] = & \
  \int_{S^2 \times M_3} d^5x\
  \text{tr} \left \langle x^M  \left| e^{-t H} \right| x^M \right\rangle   \nonumber \\ = & \
  \int_{M_3} d^3x  \
  \text{tr} \Big \langle x^\m  \Big|
  \int_{S^2} d^2 x \ \left\langle x^m \left| e^{+ t \partial_\varphi } \cdot h \cdot e^{-i \a \cdot a_0 t} \Big| x^m \right\rangle \right| x^\m \Big\rangle \nonumber \\ = & \
  \text{vol}(M_3) \times \int_{S^2} d^2 x \ \text{tr} \left\langle x^m \Big| e^{+ t \partial_\varphi } \cdot h \cdot \Big| x^m \right\rangle
\end{align}
where the symbol $\text{tr}$ indicates a matrix trace over $i$ and group indices. We used for the last
equality the fact that $a_0=0$. It works similarly for the trace of $e^{-Ht}$ over $\Psi$ fields.
One can thus reduce the index into the following form
\begin{align}
  \text{ind}\, D_{10} = \text{vol}(M_3) \times \left. \text{ind}\, D_{10}\right|_{S^2}\ ,
\end{align}
where $\left. \text{ind}\, D_{10}\right|_{S^2}$ denotes the index of the operator $D_{10}$ acting only on
the two-sphere. Since $D_{10}|_{S^2}$ is transversally elliptic, one can apply the Atiyah-Bott localization
formula to compute the reduced index:
\begin{align}
  \text{ind} \left. D_{10}\right|_{S^2} = \sum_{p:\text{fixed points}}
  \frac{\text{Tr}_{X_{S^2}(p)}\big( h  \big) -
  \text{Tr}_{\Psi_{S^2}(p)}\big(h \big)}
  {\left.\det{\big(1 -\partial \tilde x / \partial x\big)}\right|_{S^2}}\ ,
\end{align}
It is obvious to show that there are two fixed points, one of which is the north pole $\th=0$
and another is the south pole $\th=\pi$. Near the north pole, the operator $e^{Ht}$ acts
on the local coordinate $z=\th e^{i\varphi}$ as
\begin{align}
  \tilde z =q  z\ , \qquad q = e^{+it}\ .
\end{align}
It implies that
\begin{align}
  \left.\det{\big(1 -\partial \tilde x / \partial x\big)} \right|_{S^2}= ( 1- q) ( 1- q^{-1}) \ .
\end{align}
One can easily compute the value of $h$ for $X$ and $Y$ fields,
\begin{eqnarray}
  \begin{array}{rclcrcl}
    h[A_z] & = & q^{-1}, & & h[\bar \xi \g^3 \bar \l ] & = & q, \\
    h[A_{\bar z}] & = & q,  &  & h[c] & = & 1, \\
    h[\varphi_+] & = & q^{-1}, & & h[\bar c] & = & 1,
  \end{array}
\end{eqnarray}
and
\begin{eqnarray}
  \begin{array}{rclcrcl}
    h[\CA_\m] & = & 1, & & h[\bar\xi  \g_z \psi_\m - \xi \g_z \bar \psi_\m] & = & q, \\
    h[{\bar \CA}_\m] & = & 1,  &  & h[\bar\xi  \g_{\bar z} \psi_\m
  - \xi \g_{\bar z} \bar \psi_\m] & = & 1.
  \end{array}
\end{eqnarray}
%
%
Collecting all the results with the similar contribution from the south pole, one obtains
\begin{align}
  \text{ind} \left. D_{10}\right|_{S^2} = & \
  \left[- \frac{2}{1-q} + 3 \right]_{N}
  +
  \left[- \frac{2}{1-q} + 3 \right]_{S} + 2
  \nonumber \\  = & \
  \bigg[ - 2 - 2 \sum_{n=1} q^n + 3 \bigg]_N + \bigg[ 2 \sum_{n=1} (q^{-1})^n + 3 \bigg]_S + 2
  \nonumber \\ = & \
  - 2 \sum_{n=1} q^n + 2 \sum_{n=1} (q^{-1})^n + 2 \times 3\ ,
  \label{result}
\end{align}
where the first term in the braket arises from the twisted vector multiplet while the second from
twisted chiral multiplets. Finally the last term in the first line of (\ref{result}) arises from
constant modes of $\bar a_0$ and $B_0$.  One has to be careful to expand the expression in
the first line into powers in $q$.
This is due to the transversality of elliptic operators. The correct prescription is to expand the terms
from the north pole in power of $q$ and the terms from the south pole in power of $q^{-1}$.

Plugging the result (\ref{result5}) into the expression \eqref{tmpresult},
we finally obtain the following relation
\begin{align}
 Z_\text{5d SYM} = \int_{f[A_\m,\phi_\m]=0} \CD \CA_\m(x^\n) \CD \bar\CA_\m(x^\n) \
  e^{- \frac{\pi r}{g^2}\int_{M_3}\big( \CL_\text{cs}(\CA) -
  \CL_\text{cs}(\bar \CA) \big)} \ ,
\label{maintmp}
\end{align}
where the path-integral is performed over all possible field-configurations
satisfying the gauge fixing condition (\ref{f2}).
We here restore the scale of the two-sphere $S^2$.
It has turned out that
the action on the r.h.s. of \eqref{maintmp} is invariant under the
``accidental'' $G_{\mathbb{C}}$-gauge transformations,
which is larger than the original $G$-gauge transformations.
However this redundancy is completely fixed by the
gauge-fixing condition $f[A_{\mu}, \phi_{\mu}]=0$
in \eqref{f2}, making the r.h.s of \eqref{maintmp}
convergent.
This means that the relation \eqref{maintmp}
provides a path-integral definition of the
$G_{\mathbb{C}}$ Chern-Simons theory on $M_3$. \eqref{maintmp}.
%

\section{Concluding Remarks}\label{sec:remarks}

Let us now conclude with more comments on the relations \eqref{mainresult}, \eqref{trelation}.

\paragraph{Factorization}
There is a natural generalization of the relation \eqref{mainresult}.

As discussed in appendix \ref{sec:CS}, it is natural to consider the
holomorphic partition function of
the analytically continued Chern-Simons theory associated with the
classical saddle point $\alpha$ \eqref{CSblock}.
This is the basic building block for the full $G_{\mathbb{C}}$
Chern-Simons partition function \eqref{CSfactor}.

The relation \eqref{mainresult} can be thought of as a 5d lift of the
     relation
\eqref{4donI}.
 Similarly,
we expect to have a 5d lift of the relation \eqref{4donRplus}
\begin{align}
Z^{\alpha}_\textrm{5d SYM}[D \times M_3]=Z_\textrm{3d CS}^\alpha [M_3] \ ,
\label{mainfactor}
\end{align}
where $D$ is a 2d cigar (disc); a cigar asymptotes
at infinity to a cylinder, with
a $U(1)$ symmetry along the extra dimensional direction.
Reducing along this $S^1$ one obtains $D/U(1)= \mathbb{R}_{\ge 0}$,
with the tip of the cigar corresponding to the
endpoint of $\mathbb{R}_{\ge 0}$.
The label $\alpha$ represents the boundary condition of the 5d
gauge theory at infinity of the cigar.

It would be interesting to directly derive the expression (\ref{mainfactor}) from the
5d localization of the $\CN=2$ theory.
If we can show this, it follows automatically from \eqref{mainresult}, \eqref{mainfactor}
and \eqref{CSfactor} that
\begin{align}
Z_\textrm{5d SYM}[S^2 \times M_3]
=\sum_{\alpha, \bar\alpha}
n_{\alpha, \bar\alpha}\, Z_\textrm{5d SYM}^{\alpha}\left[D \times
 M_3\right]\, Z_\textrm{5d SYM}^{\bar\alpha}[\bar D \times M_3] \ ,
\end{align}
where $n_{\alpha, \bar\alpha}\in \mathbb{Z}$ and $\bar D$ is a cigar with orientation reversed.

\paragraph{6d Lift}

We expect for consistency that there are further lifts of the relations \eqref{mainresult}, \eqref{mainfactor} to 6d,
which is natural from the discussion in appendix \ref{sec:CS} ({\it cf.}
\cite{Witten:2011zz}):
\begin{align}
& Z_\textrm{6d $(2,0)$}[(S^1\times S^2)_q\times M_3]=Z_\textrm{3d CS}[M_3] \ \ ,
\label{6dlift1}
\end{align}
and
\begin{align}
& Z_\textrm{6d $(2,0)$}^{\alpha}[(S^1\times D)_q\times
 M_3]=Z_\textrm{3d CS}^{\alpha}[M_3] \ ,
\label{6dlift2}
\end{align}
where $(S^1\times S^2)_q$ and $(S^1\times D)_q$
represents twist by the $U(1)$ isometry of the $S^2$ ($D$)
along the $S^1$ direction. This twist is the 6d lift of the
similar twist for the 3d superconformal index on $S^1\times S^2$.

Since the 5d $\CN=2$ theory is believed to be equivalent with the 6d $(2,0)$ theory
on $S^2$ with finite radius $R$ (as far as the BPS sectors are concerned),
the dependence of the 6d index on twist parameter $q$ is
already captured by the dependence of the 5d partition function
on the 5d gauge coupling constant $g$.
However, for 6d intepretation we need to resum the
perturbative series expansion
in $G_{\mathbb{C}}$ Chern-Simons theory (expanded with respect to power
series in $t$)
into a power series expansion with respect to $q\sim e^{-R}\sim
e^{-g^2}$
({\it cf.}\ \cite{Kim:2012qf}).

\paragraph{General Gauge Groups}
As already stated in introduction,
our result \eqref{mainresult} strongly suggests
that the 3d/3d duality should be generalize to a general gauge group $G$
and their complexifications.
Most of the works in the literature deals with the
case $G_{\mathbb{C}}=SL(2, \mathbb{C})$.
The case of $G_{\mathbb{C}}=SL(N, \mathbb{C})$ with $N>2$
can be deal with from the mathematical work of
\cite{FockGoncharovHigher},
see also \cite{Dimofte:2013iv}  for recent discussion.
For ADE gauge groups, we expect
3d $\CN=2$ theory $\mathcal{T}_G[M]$
should arise from the boundaries of the 4d
$\CN=2$ theories of class ${\cal S}$,
similar to the cases discussed in \cite{Terashima:2011qi,Dimofte:2011ju,Cecotti:2011iy,Dimofte:2011py,Drukker:2010jp,Hosomichi:2010vh}.

\paragraph{Strong Coupling Limit}
In the relation \eqref{trelation}, the 6d limit
({\it i.e.}\ the strong coupling limit) $g\to \infty$
corresponds to the limit $t\to 0$, which
represents a highly quantum regime of
 the Chern-Simons theory \eqref{CSLag}. We can trade this with
 the classical limit $t\to \infty$ under the
 S-duality of the Chern-Simons theory $t\to t'\sim t^{-1}$, which
 originates from the S-duality of the twisted 4d $\CN=4$
 SYM.\footnote{In the notation of appendix \ref{sec:CS}, this is $\Psi\to
 \Psi'\sim \Psi^{-1}$, where the parameter $\Psi$ is the parameter for the
twisted $\CN=4$ theory and is identified with the
Chern-Simons level $t/2$.}

\section*{Acknowledgments}

We would like to thank J.~Yagi for discussion
at the initial stages of this project,
and N.~Doroud, J.~Gomis, Seok Kim, E.~Martinec, V.~Pestun and E.~Witten for
valuable discussions and helpful comments.
We would like to thank Aspen Center for Physics
(NSF Grant No.\ 1066293) for hospitality where this work has been
initiated. M.~Y.~would also like to thank Yukawa Institute for Theoretical Physics, Kyoto
University (YKIS 2012 ``Gauge/Gravity Duality'') where part of this work
has been performed. The work of S.~L.~is supported by the Ernest Rutherford fellowship of
the Science \& Technology Facilities Council ST/J003549/1.

\vskip 2cm
\centerline{\Large \bf Appendix}

\appendix

\section{3d Complex Chern-Simons}\label{sec:CS}

In this appendix we provide minimal lightening review of the
3d Chern-Simons theory with a non-compact gauge group $G_{\mathbb{C}}$ \cite{Witten:1989ip,Witten:2010cx},
and their relation with the 5d $\CN=2$ SYM.
We mostly focus on the $G_{\mathbb{C}}=SL(N, \mathbb{C})$ case.

Let us consider a 3-manifold $M_3$, and an
$SL(N, \mathbb{C})$ flat connection $\CA$
on $M_3$. We denote the complex conjugate of $\CA$
by $\bCA$.
The Lagrangian of the theory is given by
\begin{align}
\mathcal{L}_{\rm CS}[\CA, \bCA]&=\frac{t}{8\pi}
\mathcal{L}_{\rm CS}[\CA]+
\frac{\bar t}{8\pi}\mathcal{L}_{\rm CS}[\bCA] \nonumber \\
&=
\frac{t}{8\pi} \textrm{Tr}\left(
 \CA \wedge d\CA-\frac{2i}{3} \CA\wedge \CA \wedge \CA
\right) +
\frac{\bar t}{8\pi} \textrm{Tr}\left(
 \bCA \wedge d\bar\CA-\frac{2i}{3} \bCA\wedge \bCA \wedge \bCA
\right)\ ,
\label{CSLag}
\end{align}
where $t=k+i \sigma \, (k, \sigma\in \mathbb{R})$ is the complexified
level.
The parameter $k$, the ordinary level, is quantized
as usual $k\in \mathbb{Z}$ \footnote{The Lagrangian \eqref{CSLag}
relies on the choice of the trivialization of the gauge bundle,
and the partition function is independent of this choice only when
$k\in \mathbb{Z}$.
}, whereas the imaginary level $\sigma$
can be chosen to be a continuous parameter.

The partition function of the theory is defined by
\begin{align}
Z_{\textrm{$SL(N)$ CS}}[M_3]=\int \CD \CA \CD \bCA \,\,\,  e^{ i
 \int_{M_3} \mathcal{L}_{\rm
 CS}[\CA, \bCA]} \ .
\label{CSint}
\end{align}
The fact that the gauge group is non-compact means that the
trace $\textrm{Tr}$ in \eqref{CSLag} is not positive definite.
This causes a problem for the Yang-Mills kinetic terms,
making the energy unbounded from below.
However this is not problem for the Chern-Simons theory which has
a trivial Hamiltonian.

The level $t$ plays the role of the inverse Planck constant, and we can
choose to do the perturbative expansion.
The classical saddle points are given by the solutions of the equations
of motion $\CF_{\CA}=\bar \CF_{\bCA}=0$,
which is locally trivial
but global can be non-trivial due to the presence of Wilson lines.
The moduli space of the classical solutions are the moduli space of
flat connections:
\begin{align}
\mathcal{M}_{\rm flat}=\textrm{Hom}(\pi_1(M), SL(N, \mathbb{C})) \ ,
\end{align}
Let us label the holomorphic (anti-holomorphic) flat connections by
$\alpha$ ($\bar \alpha$).

Let us expand the path-integral around the classical flat connections
$\alpha, \bar\alpha$.
Since the Lagrangian \eqref{CSLag} is written as a sum of
the holomorphic and the anti-holomorphic part,
the expansion around the classical solution also factorizes
into the holomorphic part $Z_{\alpha}(t)$ and
the anti-holomorphic part $\bar Z_{\bar \alpha}(\bar t)$.
However this factorization breaks down when we choose to sum over the
flat connections:\footnote{
The equation \eqref{CSfactor} follows from the fact that the
real integration cycle of the $(G_{\mathbb{C}})_{\mathbb{C}}=
G_{\mathbb{C}}\times G_{\mathbb{C}}$ theory decomposes into
the linear combination of
$\mathcal{C}_{\alpha} \times \mathcal{C}_{\bar\alpha}$.
}
\begin{align}
Z_\textrm{CS}(t, \bar t)=\sum_{\alpha, \bar\alpha} n_{\alpha,\bar\alpha} Z^{\alpha}(t) \bar Z^{\bar \alpha}(\bar t) \ ,
\label{CSfactor}
\end{align}
for some integer coefficient $c_{\alpha, \bar\alpha}$.
The holomorphic (anti-holomorphic)
partition function $Z_{\alpha}$ ($\bar Z_{\bar\alpha}$)
can be obtained by
evaluating the path integral \eqref{CSint} over a middle-dimensional
integration cycle $\mathcal{C}_{\alpha}$ ($\mathcal{C}_{\bar\alpha}$),
defined by the downward Morse flow from the saddle point
$\alpha$ ($\bar\alpha$) with respect to the real part of the
action:\footnote{
This partition function is sometimes loosely referred to
as that of the $G_{\mathbb{C}}$ Chern-Simons theory.
However
in our terminology the $G_{\mathbb{C}}$ Chern-Simons theory
refers to the full partition function \eqref{CSfactor}, and
the holomorphic partition function \eqref{CSblock}
should rather be thought of as a analytic
continuation of the $SU(N)$ Chern-Simons theory.
}
\begin{align}
Z_{\alpha}(t)=\int_{\mathcal{C}_{\alpha}} \mathcal{D}\CA \,\,\, e^{i
 \frac{t}{8\pi} \int_{M_3} {\cal L}_{\rm CS}[\CA]} \ ,
\label{CSblock}
\end{align}

\paragraph{Relation with the 6d $(2,0)$ Theory}

Let us next quickly explain why this theory is related to
the 5d $\CN=2$ SYM discussed in this paper,
and how it is related to 3d $\CN=2$ theories ({\it cf.} \cite{Dimofte:2011py}).

Let us begin with the 6d $(2,0)$
theory on $S^1\times S^2 \times M$.
The 2-sphere $S^2$ can be thought of as
a $\tilde S^1$-fibration over an interval $I$,
where the $\tilde S^1$ fiber shrinks
at the endpoint of $I$.
By compactifying the 6d $(2,0)$ theory
on $S^1\times \tilde S^1$, we have the
the 4d $\CN=4$ theory on $I\times M$
discussed in \cite{Witten:2010cx,Witten:2010zr}.

Since $M_3$ is a curved manifold we need to topologically twist the
4d $\CN=4$ theory. The natural twist here is the one in
\cite{Marcus:1995mq,Blau:1996bx},
which is discussed in the context of geometric Langlands correspondence
\cite{Kapustin:2006pk};
this mixes the $SO(3)_R$ part of the $SO(6)_R$ R-symmetry
with the rotational $SO(3)$ of the tangential directions of $M_3$.
As explained in \cite{Kapustin:2006pk}, there is a
$\mathbb{CP}^1$-family of such topological twist, and the resulting
theory
depends only on a single parameter $\Psi$, which is a some combination
of the 4d theta-angle and the point $p$ on $\mathbb{CP}^1$.
This topological twist complexifies
the gauge field $A_{\mu}$ along $M_3$ into $\mathcal{A}_{\mu}=A_{\mu}+i
 \phi_{\mu}$,
where $\phi_{\mu}$'s are the three scalars out of the six scalars in
4d $\CN=4$ theory.
The Lagrangian of the topologically twisted theory is $Q$-exact,
except that there are boundary contributions which are given by
complex Chern-Simons terms for $\mathcal{A}_{\mu}$,
with level $t=\Psi/2$
\cite{Kapustin:2006pk}.
In our case, since we have two boundaries
we have two Chern-Simons terms, with opposite
Chern-Simons levels due to orientation reversal, leading to
\begin{align}
Z_\textrm{4d $G$ $\CN\!=\!4$}[I\times M_3]=Z_\textrm{3d $G_{\mathbb{C}}$
 CS}[M_3]\ .
\label{4donI}
\end{align}

We can instead consider the 4d $\CN=4$ SYM on
a half-line $\mathbb{R}_{\ge 0}\times M_3$.
The discussion is similar with a complex CS term
induced on the boundary,
except that we not have to specify that boundary condition
at the infinity of $\mathbb{R}_{\ge 0}$.
This is specified by a flat $SL(N)$ connection,
which we again label by $\alpha$.
\begin{align}
Z^{\alpha}_\textrm{4d $G$ $\CN\!=\!4$}[\mathbb{R}_{\ge 0}\times
 M_3]=Z^{\alpha}_\textrm{3d $G_{\mathbb{C}}$ CS}[M_3](t)\ .
\label{4donRplus}\end{align}

Lifting the story back to 5d, the interval is lifted to $S^2$
and we have the 5d $\CN=2$ SYM on $S^2\times M_3$. This means that
the result derived in the main text \eqref{mainresult}
is a 5d lift of the relation \eqref{4donI}.
In particular the fact that the two Chern-Simons terms have
opposite levels matches nicely with the result from the localization computation
\eqref{maintmp}, provided $t$ is imaginary.\footnote{This happens
when the parameter $\Psi$ is imaginary,
for example when the 4d theta-angle is trivial and the point $p$
on $\mathbb{CP}^1$ lies on the real axis.}

Similarly, we can lift the half-line $\mathbb{R}_{\ge 0}$
in \eqref{4donRplus} to a two-dimensional cigar
(disc) $D$ with a boundary condition labeled by $\alpha$,
leading to the natural relation \eqref{mainfactor}.
In this setup, the factorized form \eqref{CSfactor}
of the 3d Chern-Simons theory
is now translated into the geometrical factorization of the
$S^2$ into two cigars $D$.

Finally, we can go back to the 6d $(2,0)$ theory
on $S^2\times S^1\times M$, and choose to first compactify
on the 3-manifold $M_3$ to obtain the theory $\mathcal{T}[M_3]$
on $S^1\times S^2$. The partition function on $S^1\times S^2$,
or the 3d superconformal index, is known to
take a factorized form, which is the counterpart of
\eqref{CSfactor} \cite{Pasquetti:2011fj,Beem:2012mb}. In 3d $\CN=2$
theory the label $\alpha$ represents the vacuum of the theory
compactified on $S^1$,
and the level $t$ is related with the fugacity $q$ of the 3d index,
where the latter parameter is identified with the same parameter in the
6d $(2,0)$ theory in \eqref{6dlift1}, \eqref{6dlift2}.

\section{\texorpdfstring{5d $\CN=1$ SYM on $S^2\times \mathbb{R}^3$}{5d
N=1 SYM on S2 times R3}}\label{sec:N1SUSY}

This section is for those who are interested in the 5d SYM theory
on $S^2 \times \mathbb{R}^3$.
It turns out that one can
construct the physical (\textit{i.e.}, non-twisted) five-dimensional gauge theory with eight supercharges on
$S^2\times \mathbb{R}^3$ while preserving the full $SU(2)$ R-symmetry.
As mentioned in the beginning of section 2, the Lagrangian present below has to break
the R-symmetry $SU(2)$ completely once we replace $\mathbb{R}^3$ by $M_3$. Thus, the theory below
will not be directly related with the 3d/3d correspondence.
Nevertheless the physical theory $S^2\times \mathbb{R}^3$ itself could be of interest
from other perspectives and we report on our construction in this appendix.


\subsection{\texorpdfstring{$\CN=1$ Supersymmetry on $S^2\times
  \mathbb{R}^3$ and Lagrangian}{N=1 Supersymmetry on S2 times R3 and Lagrangian}}

\paragraph{Vector Multiplet}

Let us begin with the 5d $\CN=1$ pure SYM on $S^2\times \mathbb{R}^3$. An $\CN=1$
vector multiplet consists of a gauge field $A_M$, a real scalar field $\s$, and gaugino field
$\l_{\frak a}$ where $\frak a$ denotes $SU(2)_R$ indices.
The SUSY variation rules for the theory on the flat space $\mathbb{R}^5$ are given by
\begin{align}
  \d^{(0)} A_M & = i \e^{\frak a \frak b} \vare_{\frak a} \G_M \l_{\frak b}  \ ,
  \nonumber \\
  \d^{(0)} \sigma & =  \e^{\frak a \frak b} \vare_{\frak a} \l_{\frak b} \ ,
  \nonumber \\
  \d^{(0)} \l_{\frak a}  & = - \frac12 \G^{MN} \vare_{\frak a} F_{MN}
  + i \G^M \vare_{\frak a} D_M \sigma + \vare_{\frak b} D_{\frak a \frak c } \e^{\frak b \frak c} \ ,
  \nonumber \\
  \d^{(0)} D_{\frak a \frak b} & = - i \left( \vare_{\frak a} \G^M D_M \l_{\frak b}
  + \vare_{\frak b} \G^M D_M \l_{\frak a}  \right) +
  i \left[ \sigma, \vare_{\frak a} \l_{\frak b} + \vare_{\frak b} \l_{\frak a} \right]\ ,
\end{align}
where $D_{\frak a \frak b} = D_{\frak b \frak a}$.

In order to put the 5d $\CN=1$ pure SYM on $S^2\times \mathbb{R}^3$,
one needs to introduce additional terms to the variation rules above, coupled to the curvature of $S^2$.
The supersymmetry transformation parameters $\vare_I$ also have to satisfy a Killing spinor
equation on the two-sphere $S^2$
\begin{align}
  \nabla_m \vare_{\frak a} = \G_m \tilde \vare_{\frak a} \ , \qquad
  \partial_\mu \vare_{\frak a}  = 0
  \label{killing}
\end{align}
with
\begin{align}
  \tilde \vare_{\frak a} = - \frac{i}{2r} \G_{12} \vare_{\frak a}\ .
  \label{killing2}
\end{align}
The SUSY variation rules consistent to those in the two-sphere  $S^2$ takes the following form
\begin{align}
  \d \l_{\frak a} =  \d^{(0)} \l_{\frak a} + 2 i \sigma \tilde \vare_{\frak a} \ ,
\end{align}
while all other variations are unmodified.
One can show that this modified supersymmetry algebra closes off-shell.
The commutator $[ \d_\eta , \d_\vare] $ on the vector multiplet reads
\begin{align}
  \big[ \d_\eta , \d_\vare \big] = - i \CL_v  + \text{Lorentz}(\Th) + \text{Gauge}(\g) + SU(2)_R(\a)\ ,
\end{align}
where
\begin{align}
   \CL_v = & \ v^M \partial_M = \big( 2 \e^{\frak a\frak b} \eta_{\frak a} \G^M \vare_{\frak b} ) \partial_M\ ,
   \nonumber \\
   \g = & \ i v^M A_M + \e^{\frak a\frak b} \eta_{\frak a} \vare_{\frak b} \sigma \ ,
   \nonumber  \\
   \Th = & \   \frac{2i}{r} \e^{\frak a\frak b} \eta_{\frak a} \vare_{\frak b}\ ,
   \nonumber \\
   \a_{\frak a}^{\frak b} = & \ \frac 1r \big(\eta_{\frak a} \G_{12} \vare^{\frak b}
   + \eta^{\frak b} \G_{12} \vare_{\frak a} \big)\ .
\end{align}
The supersymmetry algebra on $S^2\times \mathbb{R}^3$ therefore becomes
a hybrid of $OSp(3|2)$ Lie superalgebra and the super Poincar\'e algebra in $\mathbb{R}^3$.

\paragraph{Supersymmetric Lagrangian}
The $\CN=1$ pure SYM Lagrangian is given by
\begin{align}
  \CL  = & \ \frac{1}{g^2} \text{tr}
  \Big[ \frac14 F_{MN}^2 +  \frac12 \big( D_M \sigma \big)^2 +  \frac i2 \e^{\frak a\frak b} \l_{\frak a}
  \G^M D_M \l_{\frak b} - \frac i2 \e^{\frak a\frak b} \l_{\frak a} \left[ \s , \l_{\frak b} \right]
  - \frac12 D_{\frak a \frak b} D^{\frak a \frak b}
  \nonumber \\ & \qquad \
  + \frac{\s^2}{2r^2}  - \frac{\s}{2r}   \e^{mn} F_{mn} \Big]
  -\frac{i}{2 r g^2}  \e^{\mu\nu\rho} \text{tr} \Big[ A_\mu \partial_\nu A_\rho - \frac{2i}{3} A_\m A_\n A_\r \Big]\ ,
  \label{N1}
\end{align}
where $r$ denotes the radius of two-sphere. Here the dimensionless
3d Chern-Simons level $k=\frac{8\pi^2 r}{g^2}$ is quantized.
Note the presence of the
the term `$\frac{\s}{r} F_{12}$', which leads to an obstruction for constructing  the physical $\CN=2$ SYM theory on $S^2\times \mathbb{R}^3$
respecting either $Sp(4)_R$ or $SO(3)_R\times SO(2)_R$ R-symmetry.

\paragraph{Comment on the Reduction to $S^2$}
In the language of 2d $\CN=(2,2)$ supersymmetry on $S^2$, {\it i.e.}\ $SU(2|1)$, the
5d $\CN=1$ pure SYM Lagrangian can be understood as a theory involving a vector multiplet
and a chiral multiplet in adjoint representation. In order to see this, one first needs
to identify $U(1)\subset SU(2)_R$ as the R-symmetry group of $SU(2|1)$.
However, this identification prevents us from performing the topological twist along $\mathbb{R}^3$.

The supersymmetry parameter $\vare_{\frak a}$ can be then decomposed as
\begin{align}
   \vare_{\frak a} = \frac{1}{\sqrt2} \Big( \xi_{a \dot a} \otimes \vare^+_{\frak a}
   - (\g^3 \bar \xi)_{a\dot a} \otimes \vare^-_{\frak a} \Big) \ ,
\end{align}
where
\begin{align}
  D_m \xi_{a\dot a} = + \frac{1}{2r} \g_m \g^3 \xi_{a \dot a} \ , \qquad
  D_m \bar \xi_{a\dot a} = - \frac{1}{2r} \g_m \g^3 \bar \xi_{a \dot a} \ .
\end{align}
and
\begin{align}
  \vare^+ = \begin{pmatrix} 1 \\ 0 \end{pmatrix}\ , \qquad
  \vare^- = \begin{pmatrix} 0 \\ 1 \end{pmatrix}\ .
\end{align}
Here $a$ denote the spinor indices on $S^2$ while $\dot a$ denotes spinor indices on $\mathbb{R}^3$.
Performing the KK reduction, one can reduce the 5d $\CN=1$ SYM down to 2d $\CN=(4,4)$ supersymmetric
theory on the two-sphere. The rotation symmetry $SO(3)_{\mathbb{R}^3}$ can be now identified as $SU(2)$ of
the $\CN=(4,4)$ supersymmetry algebra $SU(2|2)$.
Choosing
\begin{align}
  \xi_{a\dot a} = \xi_a \otimes \vare^+_{\dot a} \ , \qquad
  \bar \xi_{a \dot a} = \bar \xi_a \otimes \vare^-_{\dot a}\ ,
\end{align}
further breaks the supersymmetry down to $\CN=(2,2)$. For later convenience,
let us decompose the gaugino field $\l_{\frak a}$ into the following form
\begin{align}
  \l_{\frak a} = - \frac{1}{\sqrt 2} \Big( \l_{a} \otimes \vare^+_{\dot a} + i
  (\g^3 \psi)_a \otimes \vare^-_{\dot a} \Big) \otimes \vare^+_{\frak a}
  + \frac{1}{\sqrt 2} \Big( (\g^3 \bar \l)_a \otimes \vare^-_{\dot a}
  - i \bar \psi_a \otimes \vare^+_{\dot a} \Big) \otimes \vare^-_{\frak a}
\end{align}

Given the above reduction to 2d $\CN=(2,2)$, one can show that the above
5d $\CN=1$ SUSY transformation rules and the supersymmetric Lagrangian
can be reduced to those of a vector multiplet
$(A_m, \s, A_5, \l)$ and an adjoint chiral multiplet $(A_3,A_4,\psi)$ with R-charge $q=0$ \cite{Benini:2012ui,Doroud:2012xw}:
Upon KK reduction, it is illustrative to rewrite the bosonic part of the Lagrangian (\ref{N1})
\begin{align}
  \CL_{\rm b} = \CL_{\rm v} + \CL_{\rm c}
\end{align}
with
\begin{align}
  \CL_{\rm v}  = & \ \frac{1}{g^2} \text{tr} \Big[ \frac12 (F_{12} - \frac{\s}{r} )^2
  + (D_m \s)^2 + (D_m A_5)^2 - \big[\s, A_5\big]^2  + D^2 \Big] \ ,  \\
  \CL_{\rm c} = & \ \frac{1}{2g^2} \text{tr} \Big[ D_m \bar \phi D_m \phi
  - \big[ \s ,\bar \phi \big] \big[\s, \phi \big] - \big[ A_5 , \bar \phi \big] \big[A_5, \phi\big]
  + i \bar \phi \big[ D , \phi \big] - \frac ir \bar \phi \big[A_5 , \phi\big] + \bar F F \Big]\ ,
  \nonumber
\end{align}
where the former is the $SU(2|1)$ invariant kinetic Lagrangian for the vector multiplet while
the latter is the kinetic terms for the adjoint chiral multiplet.
It is obvious that this theory should not be embedded into the model we studied
in the main text where the latter can be reduced to $\CN=(2,2)$ twisted gauge theory on $S^2$.

\section{ \texorpdfstring{$\CN=(2,2)$ Twisted Multiplets on $S^2$}{N=(2,2) Twisted Multiplets on S2}}\label{sec:S2}

In this appendix we review the construction of
Euclidean two-dimensional  $\CN=(2,2)$
gauge theories involving a twisted vector multiplet and charged
twisted chiral multiplets studied in \cite{Gomis:2012wy,Doroud:2013pka}.

\paragraph{Twisted Vector Multiplet}
An $\CN=(2,2)$ twisted vector multiplet contains a gauge field $A_m$, two
real scalar fields $\varphi,\bar\varphi$ and gaugino field $\eta$.
The $SU(2|1)$
representation on this multiplet is given by
\begin{align}
  \d \varphi = & - \xi \g^3 \eta \ , \nonumber \\
  \d \bar \varphi = & -  \bar \xi \g^3 \bar \eta \ , \nonumber \\
  \d A_m = & \ \frac12 \big( \xi \g_m \bar \eta + \bar \xi \g^m \eta \big) \ ,\nonumber \\
  \d \eta = & \  i \g^m\g^3 D_m ( \bar \xi \varphi) + \frac i2 \xi \big[\bar \varphi, \varphi \big]
  + ( i F_{12} - D ) \g^3\xi \ , \nonumber \\
  \d \bar \eta = & \ i \g^m \g^3 D_m (\xi \bar \varphi ) + \frac i2\bar \xi \big[ \bar\varphi, \varphi\big]
  - ( i F_{12} + D) \g^3 \bar \xi \ ,   \nonumber \\
  \d D = & - \frac i2 \Big( D_m(\xi \g^m \g^3 \bar \eta + \bar \xi  \g^m \g^3 \eta ) +
  \big[ \varphi, \bar\xi \bar\eta \big] - \big[ \bar \varphi, \xi \eta
 \big] \Big) \ ,
\end{align}
The $U(1)_R$ charges of the component fields are summarized in the table below
\begin{eqnarray}
  \begin{array}{c|cccccc}
  & A_m & \varphi & \bar \varphi & \chi  & \bar \chi & D \\
  \hline
  U(1)_R & 0 & +2 & -2 & + 1 & -1 & 0
  \end{array}
\end{eqnarray}

\paragraph{Twisted Chiral Multiplet}

Let us then consider a twisted chiral multiplet in representation $\mathbf{R}$ under the
gauge group $G$. It contains a complex scalar field $Y$, a complex Dirac spinor $\chi$, and
an auxiliary field $G$. On the two-sphere, the $SU(2|1)$ supersymmetry transformation rules for
the component fields are given by
\begin{align}
  \d Y = &\  \bar \xi \g_- \chi - \xi \g_+ \chi\ , \nonumber \\
  \d \bar Y = & \ \bar \xi \g_+ \bar \chi - \xi \g_- \bar \chi \ , \nonumber \\
  \d \chi  = & \ i\g^m \g_+ \xi D_m Y - i \g^m\g_- \bar \xi D_m Y
  - i \g_- \bar \xi \varphi Y + i \g_+ \xi \bar \varphi Y \nonumber \\ &
  - \g_- \xi G + \g_+ \bar \xi G\ , \nonumber \\
  \d \bar \chi = & \ i \g^m \g_- D_m \bar Y - i \g^m \g_+ D_m \bar Y + i \g_+ \bar \xi
  \bar Y \varphi - i \g_- \xi \bar Y \bar \varphi \nonumber \\ &
  - \g_+ \xi \bar G + \g_- \bar \xi \bar G\ ,  \nonumber \\
  \d G = & \ i \xi \g_- \Big( \g^m D_m\chi - \bar \eta Y - \bar \varphi \chi \Big)
  - i \bar \xi \g_+ \Big( \g^m D_m \chi -  \eta Y - \varphi \chi \Big)\ ,
  \nonumber \\
  \d \bar G = & \ i \xi \g_+ \Big( \g^m D_m \bar \chi + \bar Y  \bar \eta + \bar \chi \bar \varphi \Big)
  - i \bar \xi \g_- \Big( \g^m D_m \bar \chi + \bar Y \eta + \bar \chi \varphi \Big) \ .
\end{align}
The $U(1)_R$ charges of the component fields are summarized in the table below
\begin{eqnarray}
  \begin{array}{c|cccccccc}
  & Y & \bar Y & \chi_- &  \chi_+  & \bar \chi_- & \bar \chi_+ & G & \bar G \\
  \hline
  U(1)_R & 0 & 0  & + 1 & -1 & -1 & +1 & 0 & 0
  \end{array}
\end{eqnarray}

\paragraph{Supersymmetric Lagrangian}
The kinetic Lagrangian for an $\CN=(2,2)$ twisted vector multiplet takes the following form
\begin{align}
  \CL_{\rm tv} = & \  \frac12 \text{tr}\Big[  F_{12}^2 + D_m \bar \varphi D_m \varphi  + \frac14
  \big[\varphi, \bar \varphi \big]^2 - i \bar \eta \g^m D_m \eta
  - \frac i2 \bar \eta \g^3 \big[\varphi, \bar \eta\big]
  + \frac i2 \eta \g^3 \big[ \bar \varphi, \eta \big]
  \nonumber \\  & \qquad + D^2 - \frac ir \bar \eta \g^3 \eta \Big] \ .
\end{align}
The $\CN=(2,2)$ twisted chiral multiplets, minimally coupled to the twisted vector multiplet,
have the kinetic Lagrangian
\begin{align}
  \CL_{\rm tc} = & \ D_m \bar Y D_m Y + \frac12 \bar Y \{ \varphi , \bar \varphi \}
  Y  + i \bar Y D Y + i \bar \chi \g^m D_m \chi+ i \bar \chi \big( \varphi \g_+ + \bar \varphi \g_- \big) \chi
  \nonumber \\ &
  + i \bar Y \big( \g_- \eta + \g_+ \bar \eta \big) \chi - i \bar \chi \big( \g_+ \eta + \g_- \bar \eta \big) Y
  + \bar G G\ ,
\end{align}
which is invariant under the $SU(2|1)$ SUSY transformation.
Twisted superpotential couplings for the twisted chiral multiplet can be written
in terms of a holomorphic function $W(Y)$. The interaction terms
\begin{align}
  \CL_W = & - i W'(Y) G - W''(Y) \chi \g_- \chi  + \frac ir W(Y) \ , \nonumber \\
  \CL_{\bar W} = & - i {\bar W}'(\bar Y) \bar G- {\bar W}''(\bar Y) \bar \chi \g_+ \bar\chi
  + \frac ir {\bar W}(\bar Y) \ ,
\end{align}
are invariant under the above $SU(2|1)$ SUSY transformation rules.

\paragraph{Remark} The Lagrangian and the supersymmetry
transformations discussed in this appendix coincide with those discussed
in section 2 when the latter are compactified along $M_3$.
The identification of various component fields are given by
\begin{gather}
  \varphi_+ = \varphi \ ,  \qquad  \varphi_- = \bar \varphi \ , \qquad \CA = Y \ , \qquad \bar \CA = \bar Y\ ,
  \nonumber \\  \l = \eta\ , \ \  \bar \l = - \bar \eta \ , \ \  \psi_+ = - \bar \chi_+ \ ,
  \ \ \psi_- = - \chi_- \ , \ \ \bar \psi_+ = -\chi_+ \ , \ \  \bar \psi_- = - \bar \chi_-  \ .
\end{gather}

\bibliographystyle{JHEP}
\bibliography{CS}

\end{document}